\documentclass[aps,11pt]{revtex4}

\usepackage[dvips]{graphics,graphicx}
\begin{document}
\title{Bloch oscillations of cold atoms in optical lattices}
\author{Andrey R. Kolovsky}
\affiliation{Max-Planck-Institut f\"ur Physik Komplexer Systeme,
                             D-01187 Dresden, Germany}
\affiliation{Kirensky Institute of Physics, 660036 Krasnoyarsk, Russia}
\email{kolovsky@mpipks-dresden.mpg.de}
\homepage{www.mpipks-dresden.mpg.de/~kolovsky}

\author{Hans J\"urgen Korsch}
\affiliation{FB Physik, University of Kaiserslautern,
                             D-67653 Kaiserslautern, Germany}
\email{korsch@physik.uni-kl.de}
\date{\today}

\begin{abstract}
This work is devoted to Bloch oscillations (BO) of cold neutral atoms
in optical lattices. After a general introduction to the
phenomenon of BO and its realization in
optical lattices, we study different extentions of this
problem, which account for recent developments in this field.
These are two-dimensional BO, decoherence of BO, and BO in
correlated systems.  Although these problems are discussed in
relation to the system of cold atoms in optical lattices,
many of the results are of general validity and can be well applied
to other systems showing the phenomenon of BO.
\end{abstract}

\maketitle

\section{Introduction}

This review is mainly addressed to researchers, who's prime scientific
interests are far from the topic announced in the title. We assume
the reader to have no preliminary knowledge of Bloch oscillations
and introduce the problem step by step, beginning from the notion of optical
lattices. On the other side, we try to avoid any extended derivations
and the theoretical analysis of the considered phenomena is presented
in a simplified form. Moreover, in some cases we only
explain the main idea of the analytical approach (referring to the
original papers) and directly proceed with the results. 
In this sense, this review serves only as an introduction to
Bloch dynamics of cold atoms. For those already familiar with 
the subject we advise to skip the first sections and move directly 
to Sec.~\ref{sec_c0} and  Sec.~\ref{sec_d0} where the most recent 
developments in the field are discussed.
  
\subsection{Brief historical review}

Originally the problem of Bloch oscillations (BO) was
formulated in context of crystalline electrons.
Considering the response of the system to a static electric
field, Zener came to the conclusion that, instead of the uniform
motion which one would naively expects, the electrons
in the crystal should oscillate \cite{Zene34}.
The period of these oscillations, now known as the Bloch
period, is given by $T_B=2\pi\hbar/dF$, where $d$ is the lattice
period and $F$ is the magnitude of the static electric force.
However, for a realistic strength of the electric field,
the period $T_B$ appears to be much smaller than the characteristic
relaxation time $\tau$ in the system. (The main contributors to 
$\tau$ are the scattering by impurities or phonons and 
the electron-electron interactions.) Because of this reason
BO have never been observed in the bulk crystal.

The status of BO as a pure theoretical problem changed after
fabrication of semiconductor superlattices \cite{Esak70}.
Here, due to the essentially larger period $d$ and the lower density
of the carriers, one can satisfy the condition $T_B<\tau$
and in 1992 the first experimental observation of BO was reported for
such systems \cite{Feld92}.
It should be stressed, however, that BO in semiconductor
superlattices are still dominated by the relaxation process.
This difficulty is overcome by optical lattices, where
standing laser waves and cold neutral atoms play the role of the
crystal lattices and the electrons, respectively.
In the latter system, the relaxation processes can be suppressed to
any desired level, which has offered unique opportunities for
experimental studies of BO \cite{Daha96,Ande98,Mors01,Cri02,Ott03,Roa04}. 

We would like also to note that the semiconductor and optical lattices
are not the only systems showing BO. As will be shown below, the
deep origin of BO lays in the band spectrum of the system. In this 
sense, any spatially periodic system may show BO. A recent
example is the periodic oscillation of a light beam in periodic
photonic structures \cite{Pert99,Sapi03}.

\subsection{List of notations}

For the sake of quick reference we list below the notations used 
throughout the paper.

\begin{itemize}

\item $\lambda$ -- the laser wave length, defines the period of 
the optical lattices $d=\lambda/2$;

\item $p_L=2\hbar k_L$ -- double recoil momentum ($k_L=2\pi/\lambda$);

\item $E_R=\hbar^2 k_L^2/2M$ -- recoil energy, defines the characteristic 
energy scale of the system;

\item $\phi_{\alpha,\kappa}(x)$ -- Bloch states, i.e. the eigenstates
of the quantum particle in a periodic potential;

\item $E_\alpha(\kappa)$ -- energy spectrum of Bloch waves, with $\alpha$ 
being the band index and $\kappa$ the quasimomentum ($-\pi/d<\kappa\le\pi/d$);

\item $\psi_{\alpha,l}(x)$ -- Wannier states, which provide an alternative
basis in the Hilbert space of the system, the site index $l$ labels
the wells of the optical lattice; 

\item $\Psi_{\alpha,l}(x)$ -- Wannier-Stark states, i.e.~the
eigenstates of the quantum particle in a periodic potential plus a
homogeneous field. Rigorously speaking, the Wannier-Stark states 
are {\em resonance} or metastable states.

\item ${\cal E}_{\alpha,l}=E_{\alpha,l} -i\Gamma_\alpha/2$ --
the spectrum of the Wannier-Stark states, where $\hbar/\Gamma_{\alpha}$
defines the lifetime of the states.

\end{itemize}

\section{Optical lattices}
\label{sec_a0}

An optical lattice is a practically perfect periodic potential
for atoms, produced by the interference of two or more laser beams.
In this section, we explain the origin of optical lattices,
their properties and some limitations.

\subsection{Optical potential}
\label{sec_a1}

The physical origin of the optical lattice is the so-called
dipole force, which acts on the atoms in the laser field. 
Indeed, let us consider a two-level atom in a standing 
plane wave:
\begin{equation}
\label{a1}
\widehat{H}=
\left( \begin{array}{cc} E_e&0\\0&E_g \end{array} \right)
+\frac{\hat{p}^2}{2M}
\left( \begin{array}{cc} 1&0\\0&1 \end{array} \right) 
-2\Omega\cos(\omega_L t)\cos(k_L x)
\left( \begin{array}{cc} 0&1\\1&0 \end{array} \right) \;.
\end{equation}
In Eq.~(\ref{a1}), $M$ is the atomic mass, $E_g$ and $E_e$ are the 
ground and excited electronic states of the atom, $\Omega$ Rabi
frequency of the dipole transition between these states,
$\omega_L$ the frequency, and $k_L$ the wave vector
of the standing wave. Substituting the wave function
$\Psi(x,t)=\exp(-i\omega_L t)\psi_e(x,t)|e\rangle+\psi_g(x,t)|g\rangle$
into the Schr\"odinger equation with the Hamiltonian (\ref{a1})
and using the rotating wave approximation, one obtains the following
system of coupled equations,
\begin{eqnarray}
\label{a2}
i\hbar\,\frac{\partial \psi_e(x,t)}{\partial t}&=&\hbar\delta\psi_e(x,t)
+\frac{\hat{p}^2}{2M}\,\psi_e(x,t) - \hbar\Omega\cos(k_L x)\psi_g(x,t) \\
\label{a3}
i\hbar\,\frac{\partial \psi_g(x,t)}{\partial t}&=&
 \frac{\hat{p}^2}{2M}\,\psi_g(x,t) - \hbar\Omega\cos(k_L x)\psi_e(x,t) \;,
\end{eqnarray}
where $\delta=(E_e-E_g)/\hbar-\omega_L$ is the detuning. Let us now 
assume that the atom is initially in its ground electronic state and
that the detuning $\delta$ is much larger than the Rabi
frequency $\Omega$. (More precisely, $\delta$ is the largest
characteristic frequency of the system.) Then 
$\psi_e(x,t)\approx(\Omega/\delta)\cos(k_L x)\psi_g(x,t)$ and we
end up with the  Schr\"odinger equation
\begin{equation}
\label{a4}
i\hbar\,\frac{\partial \psi_g(x,t)}{\partial t}=\left( 
\frac{\hat{p}^2}{2M}+V(x)\right)\psi_g(x,t) \;,
\end{equation}
\begin{equation}
\label{a5}
V(x)=V_0\cos^2(k_L x) \;,\quad V_0=-\hbar\Omega^2/\delta \;,
\end{equation}
which describes the motion of the atom along the standing wave.
The potential (\ref{a5}), which has a spatial period $d$ equal to one
half of the laser wave length, $d=\lambda/2$, is called the optical 
potential or, simply, the optical lattice. Conveniently, the depth of the 
optical lattice is measured in units of the recoil energy
$E_R=\hbar^2k_L^2/2M$. For example, for sodium atoms in a laser 
field detuned by 60 GHz from the $D_2$ sodium line (resonant wave
length 589 nm), a 4 mW power laser creates an optical potential
with an amplitude $V_0$ of about 10 recoil energies \cite{Dens02}.

\subsection{Spontaneous emission}
\label{sec_a2}

In a more sophisticated approach, where the electromagnetic field 
is treated quantum-mechanically, the dipole force appears due to
the stimulated exchange of photons between the modes of the
electromagnetic field, associated with two counter-propagating
running waves. Namely, the atom absorbs a photon from one of the running 
waves and `immediately' emits it into the other wave, getting a recoil kick 
$p_L$ along the standing wave. During this absorption-emission
process, the atom may emit a photon in the other modes of the
electromagnetic field, getting a recoil kick in an arbitrary direction.
The latter process, known as spontaneous emission (which should be
opposed to the stimulated emission discussed above) is a kind of
relaxation process due to interaction of the system with the
environment, i.e.~a bath of the electromagnetic modes. 
The rate $\tilde{\gamma}$ of spontaneous emission is given
by the product of the natural width of the excited level $\gamma$
(which is a unique characteristic of the chosen atomic transition)
and the population of the upper state,
\begin{equation}
\label{a7}
\tilde{\gamma}=\gamma (\Omega/\delta)^2 \;.
\end{equation}
This equation implies that by simultaneously
increasing the detuning and the intensity of the laser field, one 
can keep the depth of the optical potential constant but
suppresses the interaction of the system with its environment.
For example, in the case of sodium atoms in
a laser field detuned by 60 GHz, the rate of spontaneous emission 
is $\tilde{\gamma}\sim 100\;s^{-1}$, which is actually negligible 
on the time scale of the laboratory experiment \cite{Dens02}.

\subsection{Lattice dimensionality}
\label{sec_a3}

Up to now, we have considered an idealized situation of plane
waves. In practice, however, one deals with beams of a finite
width, i.e.,
\begin{equation}
\label{a8}
V({\bf r})=V_0\exp\left[-\left(\frac{r}{r_0}\right)^2\right]
\cos^2(k_L x) \;,
\end{equation}
where $r_0\sim 50\;\mu m$ is the $1/e$ diameter of the beam. 
Note that for a red detuning, the laser field also provides
a transverse confinement for the atoms. Besides the optical 
potential (\ref{a8}), there is an additional harmonic confinement
$V_{trap}({\bf r})\sim \omega_x^2 x^2+\omega_y^2 y^2+\omega_z^2 z^2$
in the laboratory experiment due to a magnetic time-orbiting potential, 
which is used to capture the atoms during the cooling procedure. 
After the sample preparation (the sample preparation includes the cooling 
of the atoms and an adiabatically switching on of the optical potential) 
this harmonic potential can be kept `switched on' or relaxed towards zero.  
If not stated otherwise, we assume the second case throughout the paper.

Using two crossed standing laser waves (4 running waves) one can
create two-dimensional lattices with approximately $(r_0/\lambda)\sim100$
wells in each direction. There is a high degree of freedom in choosing a particular 
form of the 2D potential. For example, by playing with the frequencies of 
the waves, one can realize separable `egg-crate' or non-separable
`quantum dot' potentials (see figure \ref{fig_c1} below); changing the angle between the crossing standing
waves from 90 to 60 degrees transforms the square lattice into a hexagonal 
one; and so on. We shall study 2D optical lattices in more detail 
in Sec.~\ref{sec_c0}. Needless to mention that 
using three mutually perpendicular standing waves one gets a true 
3D lattice --  an extention of the results of Sec.~\ref{sec_c0} 
to the 3D case is straightforward. 

Going ahead, we note that, beside providing a richer dynamics
of BO, the higher dimensionality of an optical
lattice also affects the strength of the atom-atom interaction.
Namely, due to a stronger confinement of the atomic wave function,
the effective constant of atom-atom interactions 
in 2D lattices is at least one order of magnitude (two - four orders for 
3D lattices) larger than in 1D lattices (see Sec.~\ref{sec_d2}).
\begin{figure}[!t]
\center
\includegraphics[width=12cm, clip]{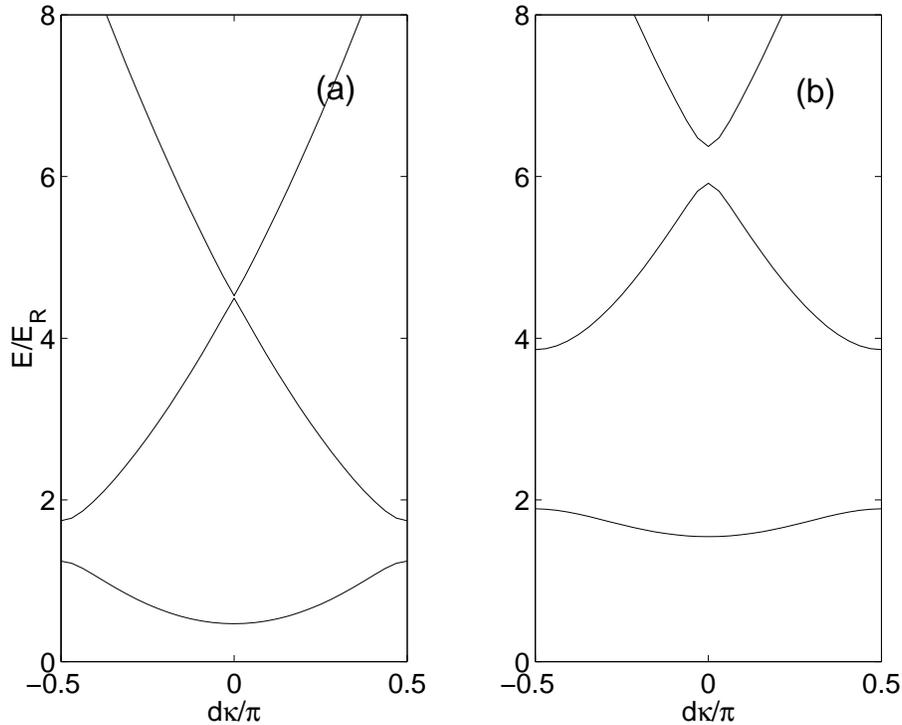}
\caption{Energy spectrum of an atom in an optical potential
of depth $V_0=E_R$ (left panel) and $V_0=4E_R$ (right panel).}
\label{fig1}
\end{figure}

\subsection{Bloch waves}
\label{sec_a4}

As well known, the energy spectrum of a quantum particle in a
periodic potential consists of the Bloch bands. An example
of the atomic Bloch band spectrum is given in Fig.~\ref{fig1},
where zero energy corresponds to the bottom of the
potential wells. If the optical lattice is switched on adiabatically,
the atoms populate the bottom of the ground band. The characteristic
width of the distribution over the quasimomentum $\kappa$ depends mainly 
on the frequency of the harmonic trap. For a small frequency (a weak 
confinement) the atoms may coherently populate hundreds of the wells,
which results in a very narrow distribution in the quasimomentum,
$\Delta\kappa\sim0.01\pi/d$. Thus one may speak of an atomic Bloch wave,
\begin{equation}
\label{a9}
\phi_{\alpha,\kappa}(x)=\exp(i\kappa x)\chi_{\alpha,\kappa}(x) 
\;,\quad \chi_{\alpha,\kappa}(x+d)=\chi_{\alpha,\kappa}(x) \;.
\end{equation}

One of the facilities provided by optical lattices is that
the atomic Bloch waves can be directly measured in the laboratory
experiment \cite{Dens02}. Usually, the measurement goes as follows.
After preparation of the Bloch wave, the optical potential
is abruptly switched off and the atoms move in free space for
a given time (the so-called `time-of-flight'). Then the atoms are
exposed to resonant light and, by `taking a picture' of the
atomic cloud, one records the spatial distribution of
the atoms. Because the time-of-flight is known, this spatial
distribution carries information about the momentum distribution
of the atoms in the optical lattice. The latter is given by the
squared Fourier transform of the Bloch wave (\ref{a9}) and
consists of a number of peaks, separated by 
$p_L=2\pi\hbar/d$. This peak-like structure of the momentum 
distribution (see Fig.~\ref{fig2} below), well observed
in the laboratory experiments, is a direct indication
of the atomic Bloch waves.

We conclude this section by introducing the {\em Wannier states}, which 
we shall use later on. These Wannier states are obtained by integrating 
the Bloch states (\ref{a9}) over the quasimomentum,
\begin{equation}
\label{a16}
\psi_{\alpha,l}(x)=\int_{-\pi/d}^{\pi/d} {\rm d}\kappa
\exp(-id\kappa l)\,\phi_{\alpha,\kappa}(x) \;,
\end{equation}
and provide an alternative basis in the Hilbert space.
Unlike the Bloch waves, the Wannier states are localized in space.
Note that, because of $\psi_{\alpha,l}(x)=\psi_{\alpha,0}(x-ld)$,
it suffices to calculate only one Wannier state ($l=0$ in what follows) 
for each energy band.

\section{Bloch oscillations in 1D lattices}
\label{sec_b0}

This section studies different regimes of BO of cold atoms 
in 1D optical lattices. It is implicitly assumed in what follows 
that neither spontaneous emission nor atom-atom interaction are important
and, thus, we can use the single-particle Schr\"odinger equation
to analyze the problem. Beside this, we assume that the transverse
motion of the atoms is frozen (i.e., we are dealing with a 1D problem).
Although we do not discuss the validity of this approximation,
the experimental results\cite{Daha96,Ande98,Dens02}
indicate that this is, indeed, the case realized in quasi 1D lattices.

\subsection{Bloch period and Landau-Zener tunneling}
\label{sec_b1}

Bloch oscillations are 
the dynamical response of the system to a static force:
\begin{equation}
\label{b10}
\widehat{H}=\widehat{H}_0+Fx \;,\quad 
\widehat{H}_0=\frac{\hat{p}^2}{2M}+V_0\cos^2(k_L x) \;.
\end{equation}
For neutral atoms, the 
static force $F$ is usually introduced by accelerating the optical 
lattice, $V(x)\rightarrow V(x-at^2/2)$, which can be done by an appropriate
chirping of the frequencies of two counter-propagating waves.
Then, in the lattice coordinate frame, the atoms experience an
inertial force of magnitude $F=aM$. The other option
is to employ the gravitational force for a vertically oriented
optical lattice, then $a=9.8\,m/s^2$. 

The common approach to the problem of BO is to look for the solution
as a superposition of Houston functions \cite{Hous40},
\begin{equation}
\label{b11}
\psi(x,t)=\sum_\alpha c_\alpha(t) \psi_\alpha(x,t) \;,
\end{equation}
\begin{equation}
\label{b12}
\psi_\alpha(x,t)
=\exp\left(-\frac{i}{\hbar}\int_0^t{\rm d}t'E_\alpha(\kappa')\right)
\phi_{\alpha,\kappa'}(x) \;,
\end{equation}
where $\phi_{\alpha,\kappa'}(x)$ is the Bloch function with  
quasimomentum $\kappa'$ evolving according to the classical equation 
of motion $\dot{p}=-F$, i.e $\kappa'=\kappa_0-Ft/\hbar$.
Substituting Eq.~(\ref{b12}) into the time-dependent Schr{\"o}dinger
equation with the Hamiltonian (\ref{b10}), we obtain
\begin{equation}
\label{b13}
i\hbar\dot{c}_\alpha=F\sum_\beta X_{\alpha,\beta}(\kappa')
\exp\left(-\frac{i}{\hbar}\int_0^t{\rm d}t'[E_\alpha(\kappa')
-E_\beta(\kappa')]\right) c_\beta \;,
\end{equation}
where $X_{\alpha,\beta}(\kappa)=\int{\rm d}x 
\chi^*_{\alpha,\kappa}(x)\partial\chi_{\beta,\kappa}(x)/\partial\kappa$.
When neglecting the inter-band coupling, i.e $X_{\alpha,\beta}(\kappa)=0$
for $\alpha\ne\beta$, we have
$i\hbar\dot{c}_\alpha=F X_{\alpha,\alpha}(\kappa')c_\alpha$ 
and, thus,
\begin{equation}
\label{b14}
|c_\alpha(t)|=|c_\alpha(0)| \;.
\end{equation}
This solution is the essence of the so-called {\em single-band
approximation}.
\begin{figure}[!t]
\center
\includegraphics[width=12cm, clip]{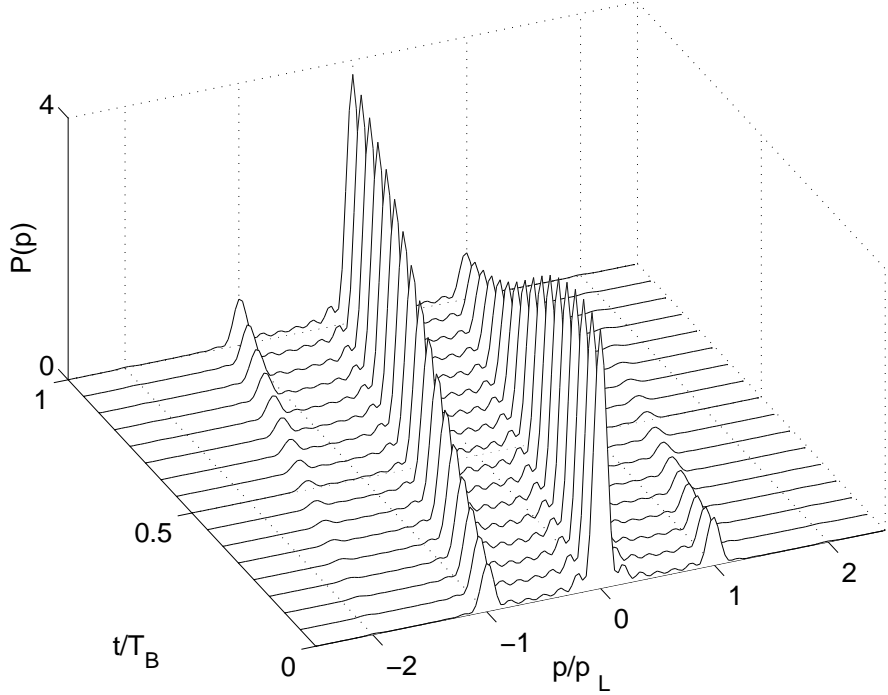}
\caption{Dynamics of the atomic momentum distribution $P(p)$, induced by
a weak static force $F<0$. The depth of the optical potential is
$V_0=10E_R$. The weak wiggling of $P(p)$ is an artifact due to
the finite size of the lattice ($L=10$, periodic boundary conditions)
used in the numerical simulations.}
\label{fig2}
\end{figure}

The correction to the solution (\ref{b14}) is obtained by using the
formalism of Landau-Zener tunneling. In fact, when the quasimomentum 
$\kappa'$ explores the Brillouin zone, an adiabatic transition occurs at
the points of `avoided' crossings between the adjacent Bloch bands 
(see, for example, the avoided crossing between the 1st and 2nd band 
in Fig.~\ref{fig1}(a) at $\kappa=\pi/d$). As a result, the population
of the $\alpha$th band decreases exponentially with the decay time
\begin{equation}
\label{b15}
\tau=\hbar/\Gamma_\alpha \;,\quad
\Gamma_\alpha=a_\alpha F\exp(-b_\alpha/F) \;,
\end{equation}
where $a_\alpha$ and $b_\alpha$ are band-dependent constants.
Note, that Eq.~(\ref{b15}) provides only an estimate 
for the mean decay rate and, to find the exact dependence
$\Gamma_\alpha(F)$, one has to employ a different approach
which we shall briefly discuss later on in Sec.~\ref{sec_b4}.

As follows from the estimate (\ref{b15}) for a weak static force
the Landau-Zener tunneling can be neglected and the solution of the 
problem is essentially given by Eq.~(\ref{b12}), i.e., 
$\psi(x,t)\sim \phi_{\alpha,\kappa_0-Ft/\hbar}(x)$.
The linear change of the quasimomentum of the Bloch wave results in 
a periodic change of the atomic momentum distribution $P(p)=|\psi(p,t)|^2$
(see Fig.~\ref{fig2}) which is the quantity measured in 
in the laboratory experiments \cite{Daha96,Roa04}. (Since the measurement 
is destructive, one has to repeat the experiment several times to record 
the time-evolution of the momentum distribution.)
The period of these oscillations is given by the Bloch period,
\begin{equation}
T_B=2\pi \hbar/d F 
\end{equation}
with $d=\lambda/2$ and $F=aM$. Using the single-band approximation,
it is also easy to show that the mean momentum evolves as  
$\langle p(t)\rangle=M v(Ft)$, where 
$v(\kappa)=\partial E_\alpha(\kappa)/\hbar\partial\kappa$ is the
group velocity. Note, that the amplitude of oscillations of 
$\langle p(t)\rangle$ is proportional to the band width and, thus, 
can be extremely small (deep optical lattices). The oscillations of
the atomic momentum distribution, however, are qualitatively
the same independent of the particular choice of the parameters and, 
in this sense, are a more reliable signature of BO.

\subsection{Wannier-Stark ladder}
\label{sec_b2}

Bloch oscillations can also be described
in terms of the Wannier-Stark states, which provide a
useful insight into the physics of this phenomenon. 
To simplify the analysis, we consider a tight-binding model -- 
an additional (after the single-band) approximation
to the original problem. This approximation is not crucial
and one obtains similar results for the single-band model.
Using the notion of the Wannier states (\ref{a16}),
the Hamiltonian of the tight-binding model has the form,
\begin{equation}
\label{b17}
\widehat{H}_{TB}=\sum_l(E_\alpha+dFl)\,|l\rangle\langle l|
+\frac{J_\alpha}{2}\sum_l
\big(\,|l+1\rangle\langle l|+|l-1\rangle\langle l|\,\big) \;,
\end{equation}
where $J_\alpha$ is the hopping matrix element, $E_\alpha=
\overline{E_\alpha(\kappa)}$, and $\langle x|l\rangle=
\psi_{\alpha,l}(x)$. The Hamiltonian (\ref{b17}) can be 
easily diagonalized, giving the spectrum,
\begin{equation}
\label{b18}
E_\alpha(\kappa)=E_\alpha+J_\alpha\cos(d\kappa) \;,
\quad {\rm if} \quad F=0 
\end{equation}
(note, in passing, that the tight-binding model approximates the Bloch 
dispersion relation by a cosine function), and
\begin{equation}
\label{b19}
E_{\alpha,l}=E_\alpha + dFl \;,\quad 
{\rm if} \quad F\ne0 \;.
\end{equation}
The discrete spectrum (\ref{b19}) 
is known as the {\em Wannier-Stark ladder} and the eigenfunctions 
corresponding to $E_{\alpha,l}$,
\begin{equation}
\label{b20}
|\Psi_{\alpha,l}\rangle
=\sum_m {\cal J}_{m-l}\left(\frac{J_\alpha}{2dF}\right)|m\rangle \;,
\end{equation}
(here ${\cal J}_n(z)$ is the ordinary Bessel function) are known
as the {\em Wannier-Stark states}. Because the Bessel functions 
${\cal J}_n(z)$ are exponentially small for $|n|>|z|$, the
Wannier-Stark states are localized in space with a
localization length $l_{WS}=1$ (in units of lattice period)
for $dF>J_\alpha$ and $l_{WS}\approx J_\alpha/dF$ for $dF<J_\alpha$. 
One may refer to these two cases as the `strong force' and `weak force'
regimes. In this paper, however, we reserve the term
`strong force' to static force magnitudes which break the 
single-band approximation (strong Landau-Zener tunneling).

\subsection{Wave packet dynamics}
\label{sec_b3}

\begin{figure}[!t]
\center   
\includegraphics[width=12cm, clip]{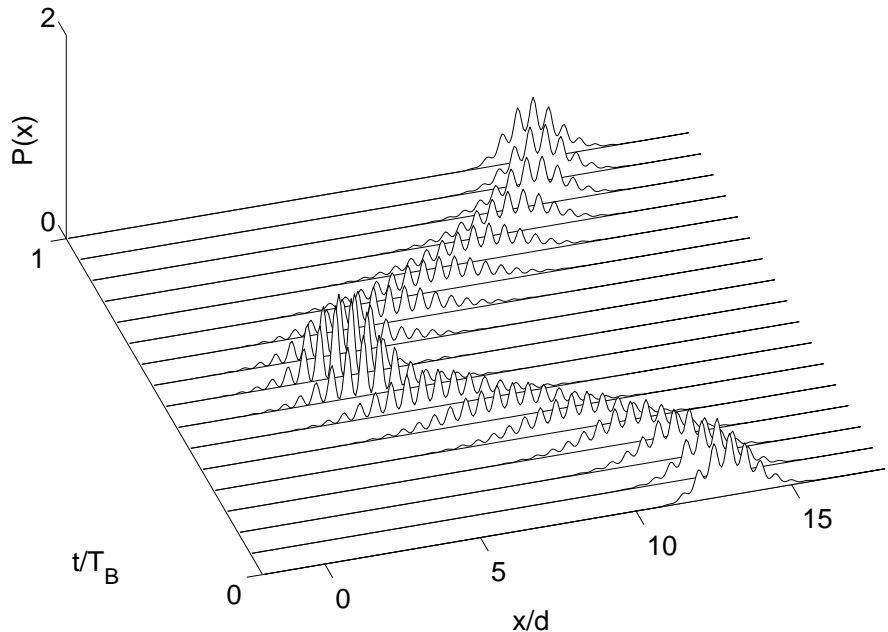}
\caption{Spatial oscillations of the localized wave-packet.
The amplitude of the oscillations is given by the localization
length of the Wannier-Stark states $l_{WS}\sim1/F$.}    
\label{fig3}   
\end{figure}

In Sec.~\ref{sec_b1} we have considered the case of a Bloch wave 
as an initial condition. It is also interesting to study a situation, 
where only a few wells of the optical potential 
are populated \cite{Hart03}. Then, along with the oscillations in
momentum space, the atoms also oscillate in configuration
space. This is illustrated in Fig.~\ref{fig3} which shows
the dynamics of $P(x)=|\psi(x,t)|^2$ for a `minimum uncertainty'
wave packet. The amplitude of the oscillations is
given by the localization length of the Wannier-Stark states.
Indeed, since the general solution of the Schr\"odinger equation
can be written as a sum over the Wannier-Stark states,
\begin{equation}
\label{b21}
\psi(x,t)=\sum_l c_l\exp(-iE_{\alpha,l}t/\hbar)\,\Psi_{\alpha,l}(x) \;,
\end{equation}
the localization length $l_{SW}$ defines the maximum distance
where the wave-packet can move to. 

For the purpose of future use, we also display the solution
(\ref{b21}) in the momentum representation. Expanding the initial
state in terms of Bloch states, $\psi(x,0)=\int d\kappa 
g_\alpha(\hbar\kappa)\phi_{\alpha,\kappa}(x)$, we introduce the 
envelope functions $g_\alpha(p)=g_\alpha(\hbar\kappa)$.
Then, using the extended Brillouin zone representation, 
the solution $\psi(p,t)$ can be represented as
\begin{equation}
\label{b22}
\psi(p,t)=\sum_\alpha \exp(-iE_{\alpha}t/\hbar)
\sum_{n=-\infty}^{\infty}g_\alpha(p+Ft+np_L)\,\Psi_{\alpha,0}(p)
\end{equation}
(in comparison with Eq.~(\ref{b21}) here we also included the sum of 
$\alpha$). Equation (\ref{b22}) is well suited for a numerical simulation
of Bloch dynamics and has actually been used in Sec.~\ref{sec_b1}
to illustrate the oscillations of the momentum distribution.

\subsection{Bloch oscillations for strong static forcing}
\label{sec_b4}

Above we have considered only the case of a weak static force, where 
the Landau-Zener tunneling is negligible and one can use the
single-band approximation to study the system dynamics.
This section is devoted to the regime of strong forcing,
which we shall analyze by using the formalism of {\em metastable} 
(or resonance) Wannier-Stark states. The latter are defined as 
non-hermitian eigenstates of the Hamiltonian (\ref{b10}), 
corresponding to the complex energies
\begin{equation}
\label{b23}
{\cal E}_{\alpha,l} = E_\alpha + dFl-i\Gamma_\alpha/2 \;.
\end{equation}
The spectrum (\ref{b23}) generalizes the notion of the Wannier-Stark 
ladder (\ref{b19}) and is given by the poles of the scattering matrix
of the system. For the details of the scattering matrix approach to 
the Wannier-Stark problem we refer the reader to the review \cite{pr53}.
Here we only note that this approach does not involve any approximation
and, hence, the results presented below are rigorous.

\begin{figure}[!t]
\center
\includegraphics[width=12cm, clip]{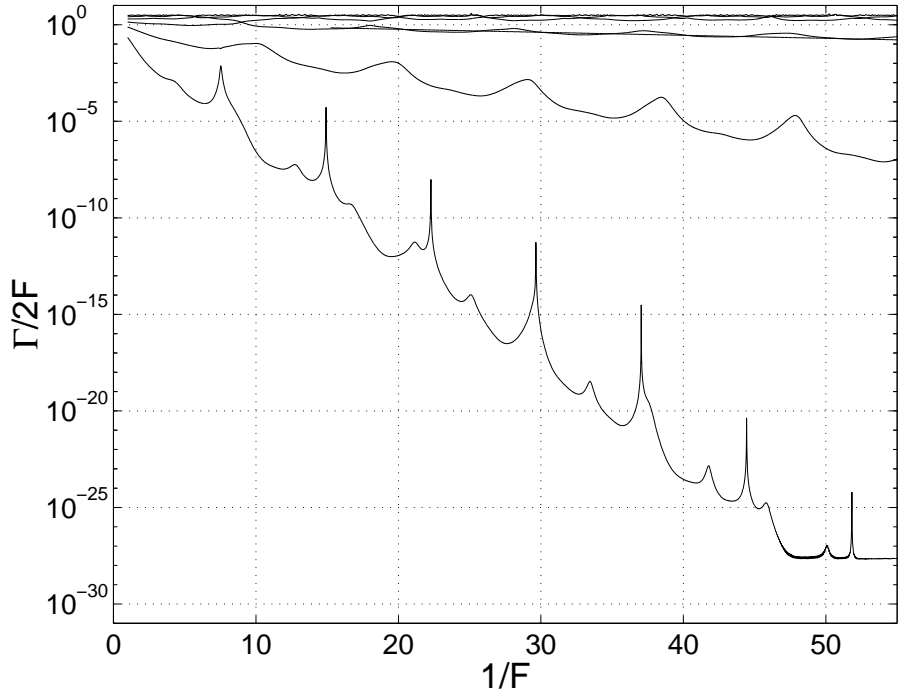}
\caption{Imaginary part of the complex energies (\ref{b23})
as the function of the inverse scaled static force $1/F$
($F\rightarrow dF/2\pi V_0$). The depth of the optical
potential is $V_0=8E_R$.}
\label{fig4}
\end{figure}

The decay constant $\Gamma_\alpha$ in Eq.~(\ref{b23}), which
defines the lifetime of the Wannier-Stark states, has a rather
nontrivial dependence on the static force.  An example is given
in Fig.~\ref{fig4}, where the six different curves correspond
to $\Gamma_\alpha$ of the Wannier-Stark ladder, originating from the
six lowest bands of the Hamiltonian $H_0$. 
Strong fluctuations of the decay rate, superimposed on the Landau-Zener 
dependence (\ref{b15}), are noticed. These fluctuations are due to resonance
tunneling, occurring when the positions of the Wannier-Stark levels 
in different wells of the optical potential coincide, i.e.,
when ${\rm Re}[{\cal E}_{\alpha,l}(F)]={\rm Re}[{\cal E}_{\beta,m}(F)]$.

Having the scattering problem solved, one generalizes Eq.~(\ref{b22})
for the system dynamics simply by substituting the
stationary Wannier-Stark states (\ref{b20}) by the metastable
Wannier-Stark states, the real energy $E_\alpha$ by the
complex energy ${\cal E}_\alpha=E_\alpha-i\Gamma_\alpha/2$,
and multiplying the whole expression by the Heaviside step-function
$\Theta(p+Ft)$ which truncates the momentum distribution 
at $p<-Ft$ \cite{pr53}:
\begin{equation}
\label{b22a}
\psi(p,t)=\Theta(p+Ft)\sum_\alpha\exp(-iE_{\alpha}t/\hbar)
\sum_{n=-\infty}^{\infty}g_\alpha(p+Ft+np_L)\,\Psi_{\alpha,0}(p) \;. 
\end{equation}
The characteristic feature of the resonance Wanier-Stark states 
$\Psi_{\alpha,0}$ is an exponentially growing tail for negative
momentum, $\Psi_{\alpha,l}(p)\sim\exp(\Gamma_\alpha p/\hbar F)$.
Then, as follows from Eq.~(\ref{b22a}),
the solution $\psi(p,t)$ is essentially a sequence of
wave packets, separated by a distance $p_L$.
In the coordinate representation, these equidistant
sequence of the `momentum' wave packets transforms into a train 
of `coordinate' wave packets, distributed in space according to 
a square law. Thus, in the strong field
regime, the atomic array acts as a matter laser, emitting one
pulse of matter per Bloch period \cite{Ande98}. 

\subsection{Related problems}
\label{sec_b5}

In this subsection, we briefly discuss two important modifications 
of the problem of atomic BO. The first one deals with BO in the 
presence of harmonic confinement \cite{Thom03}, i.e.
$V(x)=V_0\cos^2(k_L x)+M\omega_x^2 x^2/2$. (The characteristic
value of the frequency $\omega_x$ is a few Hz, which should be
compared with the frequency of small atomic oscillations  
$\sim \omega_R(V_0/E_R)^{1/2}$ of few kHz.)
Let us consider the situation where the atoms are located 
far from the trap origin. In practice, this initial condition
is realized by a sudden shifting
of the center of the trap to a distance $x_0\gg d$ \cite{Ott03}.
Then the atoms locally feel a static force $F=M\omega_x^2 x_0$,
which is one of the preconditions for BO. It should be noted, however,
that the analogy with BO should be drawn here with some precautions.
Indeed, using the tight-binding approximation, the Hamiltonian
of the atom in the combined potential reads as
\begin{equation}
\label{b24}
\widehat{H}_{eff}=\sum_l \frac{\nu}{2} l^2|l\rangle\langle l|
+\frac{J_\alpha}{2}\sum_l \big(\,|l+1\rangle\langle l|
+|l-1\rangle\langle l|\,\big)
\end{equation}
with $\nu=M\omega_x^2 d^2$. This Hamiltonian formally corresponds
to the Hamiltonian of the quantum pendulum,
\begin{equation}
\label{b25}
\widehat{H}_{eff}=-\frac{\nu}{2}\frac{d^2}{d\theta^2}+J_\alpha\cos\theta \;,
\end{equation}
and, hence, the characteristic frequency of atomic oscillations is 
given by the frequency of the pendulum. The latter is known to
have a rather nontrivial dependence on pendulum energy and 
can be approximated by a linear law only in the asymptotic
region of large energies \cite{Lich83}. 
Thus, to mimic BO (in a sense that the frequency of
oscillations is inversely proportional to the local static field)
one should satisfy the condition that the pendulum (\ref{b25})
is well above its separatrix. 

The second problem we would like to mention deals with 
BO in a resonant or near-resonant laser field.
In this case, the external
degree of freedom of the atom (the motion of the center
of mass) cannot be decoupled from its internal (electronic)
degree of freedom and, thus, we have to solve the
system of  partial differential equations (\ref{a2}-\ref{a3}) 
(with a static term added) exactly, without adiabatic elimination
of the upper state \cite{job54}. It was found,
in particular, that BO of the atoms in a resonant field
enforce a kind of Rabi oscillations, where up to 90 percent of
the atoms may appear in the excited electronic state. We note,
however, that an experimental realization of this interesting regime 
of BO requires a very narrow width of the optical transition
$\hbar\gamma<E_R$.

\section{Bloch oscillations in 2D lattices}
\label{sec_c0}

In this section, we study Bloch oscillations in 2D optical
lattices. We restrict ourselves to square lattices, created
by laser beams of equal intensities. If the frequencies
of the crossing standing waves also coincide, the optical potential
is given by 
\begin{equation}
\label{c1}
V(x,y)=V_0[\cos(k_Lx)+\cos(k_Ly)]^2 \;, 
\end{equation}
as can be easily shown by repeating the derivation of Sec.~2.1. Note that
the potential (\ref{c1}) is not separable. If, however, 
the frequencies of the waves are slightly mismatched, we obtain 
a separable potential 
\begin{equation}
\label{c2}
V(x,y)=V_0\big[\,\cos^2(k_Lx)+\cos^2(k_Ly)\,\big] \;.
\end{equation}
Obviously, the property of separability can be attributed only to
2D or 3D potentials. The consequences of this property for the
dynamics of BO is one of the main questions we address in this section.
\begin{figure}[!h]   
\center
\includegraphics[width=8cm, clip]{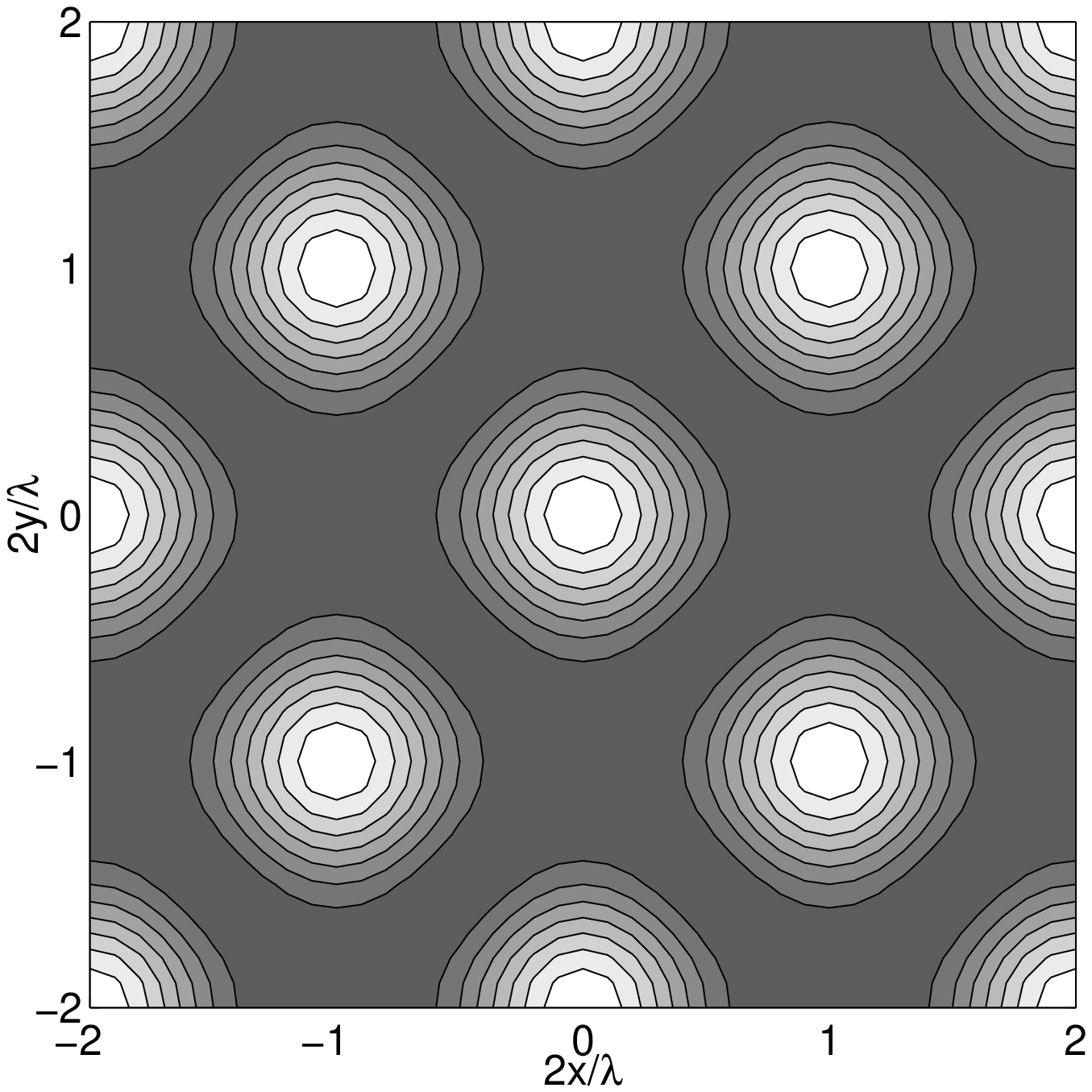}
\includegraphics[width=8cm, clip]{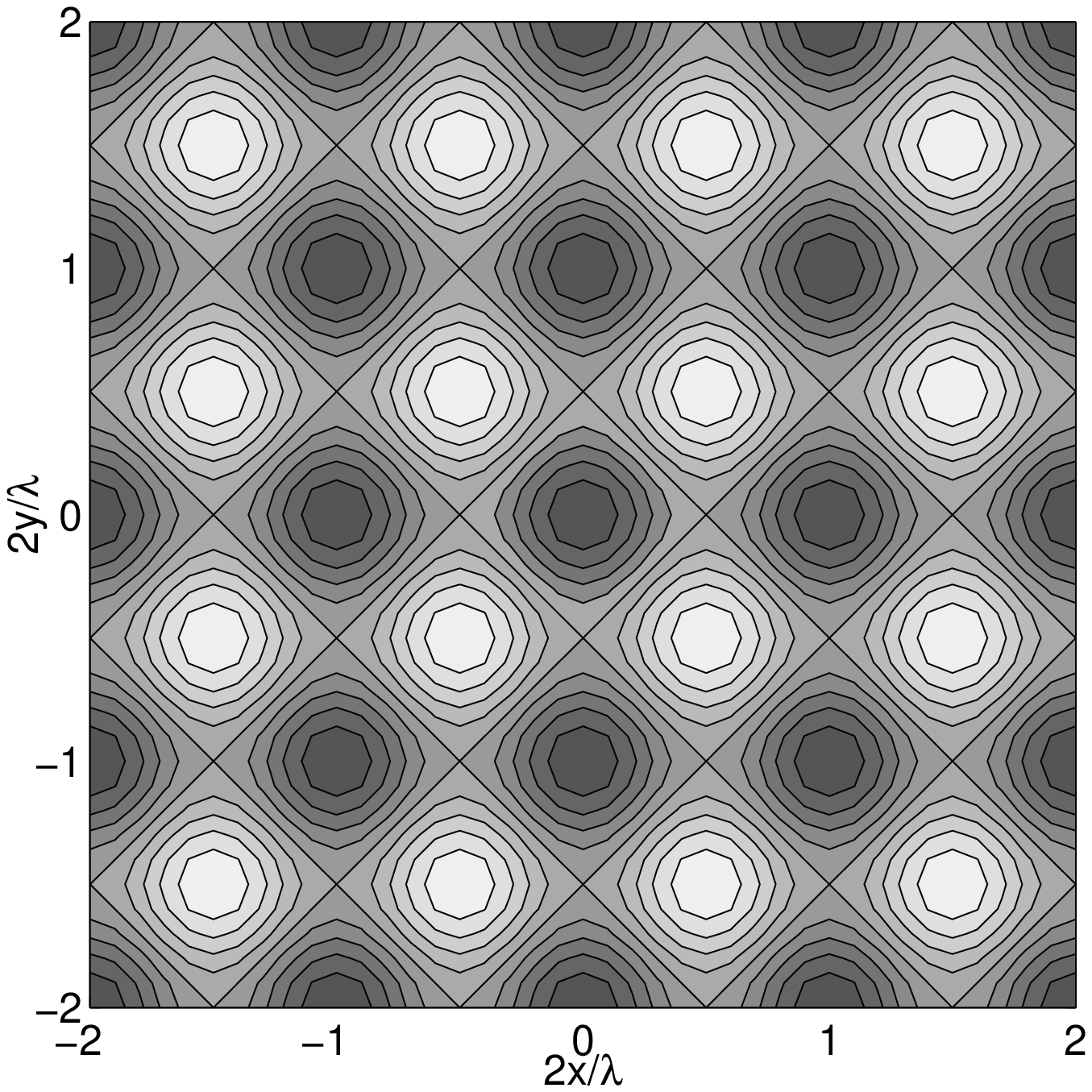}
\caption{Contour plot of the non-separable optical potential (\ref{c1}) 
(left panel), and the separable potential (\ref{c2}) (right panel).
Note that the primitive translation vectors for the lattice (\ref{c1})
are rotated by 45 degrees with respect to the laboratory coordinate
system.}
\label{fig_c1}
\end{figure}

\subsection{Band spectrum}
\label{sec_c1}

We begin with the analysis of the Bloch band spectrum
for the specified 2D lattices. To simplify the equations, we shall use 
a scaling where the coordinate is measured in units of 
the lattice period $d$ and the time in periods of the recoil 
frequency ($t\rightarrow \omega_R t$). This scaling (which also
involves a rotation of the coordinate system) leads to the 
dimensionless Hamiltonian
\begin{equation}
\label{c3}
\widehat{H}_0=-\frac{\hbar^2}{2}\left(\frac{d^2}{dx^2}
+\frac{d^2}{dy^2}\right) +\cos x + \cos y -\varepsilon\cos x\cos y 
\end{equation}
where the only independent parameter is the the scaled Planck's constant 
$\hbar\sim(E_R/V_0)^{1/2}$ and the constant $\varepsilon$ equal to zero 
or $\pm1$ for a separable and non-separable potential, respectively.
(Although $\varepsilon$ can take only the specified integer values, for 
theoretical purposes it might be useful to consider the whole interval 
$0\le|\varepsilon|\le1$.)

\begin{figure}[!t]   
\center
\includegraphics[width=12cm, clip]{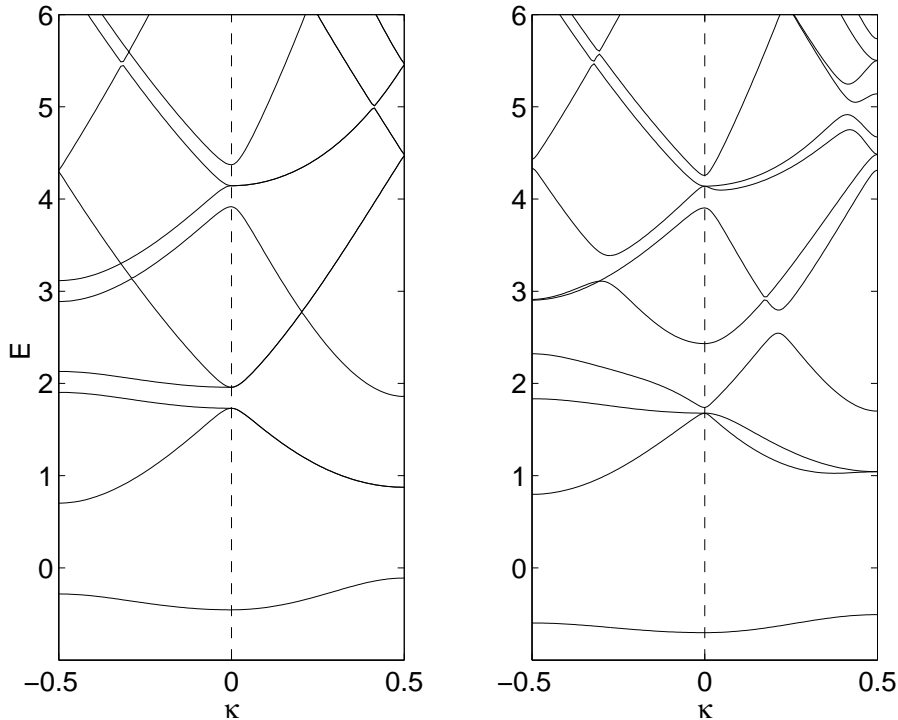}
\caption{Cross-section of the energy spectrum $E_\alpha(\kappa_x,\kappa_y)$
along the lines $\kappa_y=0$ ($\kappa_x<0$) and
$\kappa_y=\kappa_x$ ($\kappa_x>0$) for $\varepsilon=0$ (left panel)
and $\varepsilon=1$ (right panel). The value of the scaled Planck's
constant is $\hbar=2$.}
\label{fig6}
\end{figure}

To characterize the two-dimensional dispersion relation
$E_\alpha(\kappa_x,\kappa_y)$, we consider the cross-section of
the spectrum along the lines $\kappa_y=0$ for $\kappa_x<0$ and
$\kappa_y=\kappa_x$ for $\kappa_x>0$. The result is depicted in
Fig.~\ref{fig6}, where the left and right panels refer to the
case $\varepsilon=0$ and $\varepsilon=1$, respectively. It is seen
in the figure that a nonzero value of $\varepsilon$ strongly modifies the
central part of the spectrum. (In particular, it removes the
degeneracy between the bands along the lines $\kappa_y=\pm\kappa_x$.) 
As concerning the ground Bloch band, the difference shows up in the 
coefficients of the Fourier expansion of the dispersion relation
\begin{equation}
\label{c4}
E_0(\kappa_x,\kappa_y)
     =\sum_{m,n}J_{m,n}\exp(i2\pi m\kappa_x)\exp(i2\pi n\kappa_y) \,.
\end{equation}
In the separable case, only the coefficients $J_{m,n}$ with $n=0$ or $m=0$ 
differ from zero while in the non-separable case all 
elements have non-zero values. (Also note the symmetry $J_{\pm n,\pm m}
=J_{m,n}$.) At the same time, even the largest non-trivial 
coefficient $J_{1,1}$ is only $1/20$ of the coefficient $J_{0,1}$, 
which practically alone determines the width of the ground Bloch band.

\subsection{Fractional Wannier-Stark ladder} 

Let us briefly discuss the energy spectrum of the 2D 
Wannier-Stark system,
\begin{equation}
\label{c5}
\widehat{H}=\widehat{H}_0+F_x x+ F_y y \,, 
\end{equation}
where $\widehat{H}_0$ is given in Eq.~(\ref{c3}). First we consider 
the special case when the vector ${\bf F}$ of the static force is parallel
to one of the crystallographic axes of the lattice (the $x$-axis, to be
certain). In this case, the Hamiltonian (\ref{c5}) possesses the
`ladder symmetry' along the field direction and a translational
symmetry in the direction perpendicular to the field. Thus, the
spectrum of the system consists of replica of the Bloch
band ${\cal E}_\alpha(\kappa_y)$, shifted relative to each other
by the Stark energy $2\pi F$ (here $2\pi$ stands for the lattice period). 

The above result can be  extended to the case of arbitrary 
`rational' directions of the field,
\begin{equation}
\label{c6}
\frac{F_x}{F_y}=\frac{q}{r} \;,
\end{equation}
where $q,r$ are co-prime integers. In this case, one uses
the transformation of the coordinates \cite{Naka93},
\begin{equation}
\label{c7}
x'=\frac{qx+ry}{(r^2+q^2)^{1/2}} \;,\quad 
y'=\frac{qy-rx}{(r^2+q^2)^{1/2}}\;,
\end{equation}
which introduces a new lattice with the period $d'=2\pi(r^2+q^2)^{1/2}$
and matches the vector ${\bf F}$ to the primitive vector
of this new lattice,
\begin{equation}
\label{c8} 
\widehat{H}'=\frac{\hat{p}^{'\,2}_x}{2}+\frac{\hat{p}^{'\,2}_y}{2}+V(x',y')+Fx' \;.
\end{equation}
It is also easy to see that the transformation (\ref{c7}) actually
introduces $s=r^2+q^2$ different (sub)lattices, whose Hamiltonians
differs from (\ref{c8}) by an additive term $(d'F/s)j$, $j=1,\ldots,s-1$.
Thus, for rational directions of ${\bf F}$, the spectrum of the 
original Hamiltonian (\ref{c5}) is a fractional Wannier-Stark ladder 
along the direction of the field, constructed from the Bloch bands 
with $\sqrt{s}$ times reduced Brillouin zone in the direction
perpendicular to the static force, i.e.,
\begin{equation}
\label{c9}
{\cal E}_{\alpha,l}(\kappa_\perp)={\cal E}_{\alpha}(\kappa_\perp)
+\frac{2\pi Fl}{(r^2+q^2)^{1/2}} \;, \quad
{\cal E}_{\alpha}(\kappa_\perp)=E_{\alpha}(\kappa_\perp)
-i\frac{\Gamma_\alpha(\kappa_\perp)}{2} \;.
\end{equation}
Note that for a separable potential the sub-bands 
${\cal E}_{\alpha}(\kappa_\perp)$, $-s^{-1/2}\le \kappa_\perp <s^{-1/2}$,
have zero width for any direction of
the fields $\theta=\arctan(r/q)$, except $\theta=0,\pi/2$.
(For the imaginary part of the dispersion relation one obviously
has $\Gamma_\alpha(F,\theta)=\widetilde{\Gamma}_\alpha(F\cos\theta)+
\widetilde{\Gamma}_\alpha(F\sin\theta)$, where $\widetilde{\Gamma}_\alpha(F)$
is obtained by solving the 1D problem.) As $\varepsilon$
deviates from zero, the sub-bands gain a small but finite width. 
A more dramatic consequence of the non-separability, however,
is the strong dependence of the decay rate $\Gamma_\alpha(\kappa_\perp)$
on the quasimomentum $\kappa_\perp$, where the decay rate may
vary by several orders of magnitude \cite{prl51}.

\subsection{Wave packet dynamics}
\label{sec_c3}

It is interesting to compare the wave packet dynamics for
separable and non-separable potentials. Restricting ourselves
to the ground Bloch band, the results of these studies \cite{pra58,Witt04}
can be summarized as follows.

For a weak static force (negligible Landau-Zener tunneling) and 
$\varepsilon=0$, the two-dimensional BO are given by a superposition 
of the one-dimensional BO. In other words, BO is a (quasi)periodic
process with two periods defined by the projections of the static force 
to the crystallographic axes of the lattice,
$T_{x,y}=2\pi\hbar/dF_{x,y}$. In coordinate space, this is reflected in
Lissajous-like trajectories of a localized wave packet in the $xy$-plane. 
Note that for $\varepsilon=0$, the motion of the wave packet is 
generally non-dispersive. (Exclusions are $\theta=0,\pi/2$, where
BO along one axis are accompanied by a dispersive spreading  of the
wave packet along the other axis.) A nonzero $|\varepsilon|\le1$ only 
slightly modifies this dynamics which is actually not surprising, because
the dispersion relation (\ref{c4}) for the ground Bloch band can
be well approximated by a `separable' dispersive relation,
$E(\kappa_x,\kappa_y)\approx(J_{1,0}/2)\cos(2\pi\kappa_x)+
(J_{0,1}/2)\cos(2\pi\kappa_y)$. In terms of the energy spectrum
(\ref{c9}) this approximation amounts to neglecting the width
of the bands $E_{\alpha}(\kappa_\perp)$.
\begin{figure}[!t]   
\includegraphics[width=8cm, clip]{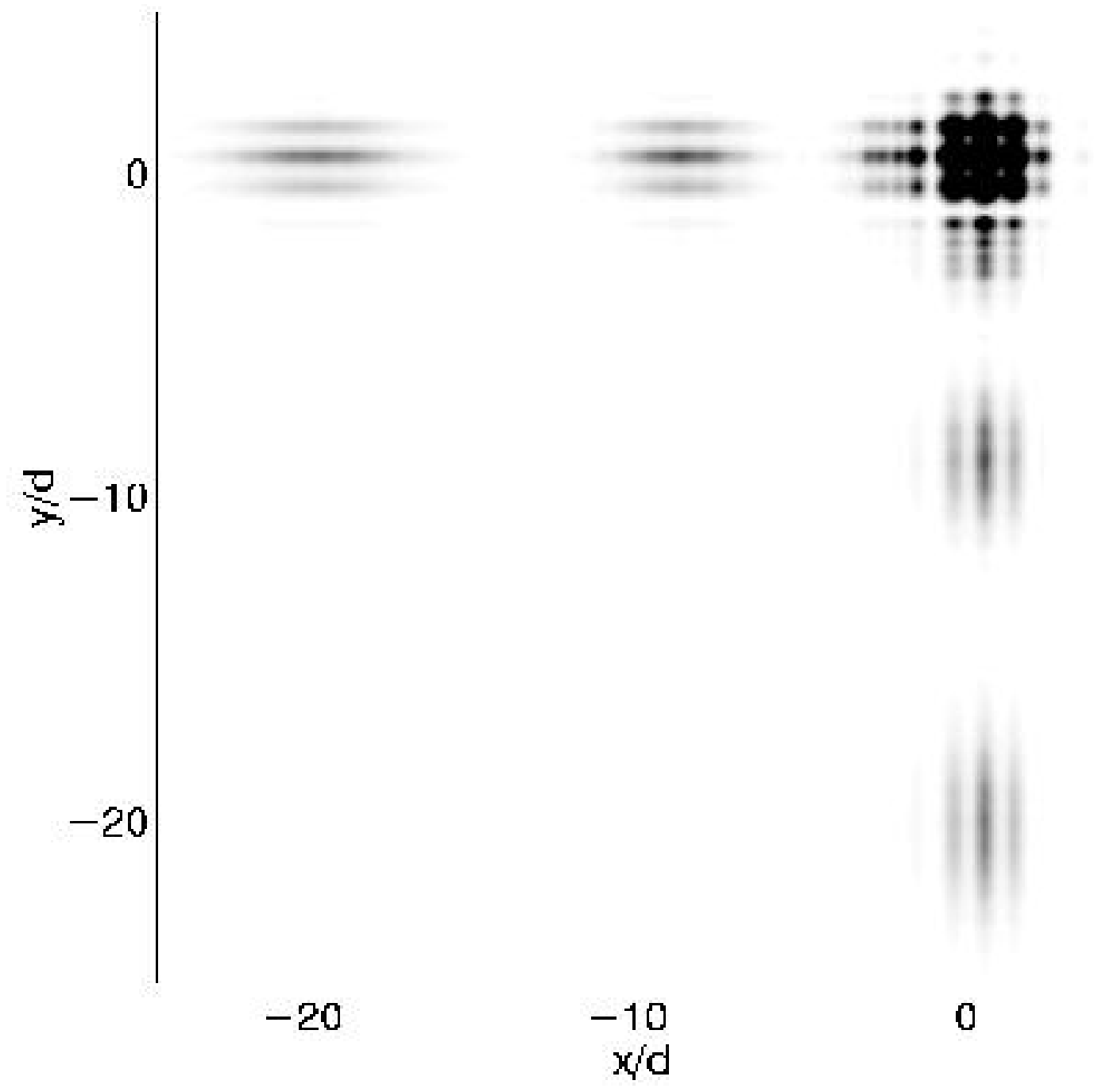}
\includegraphics[width=8cm, clip]{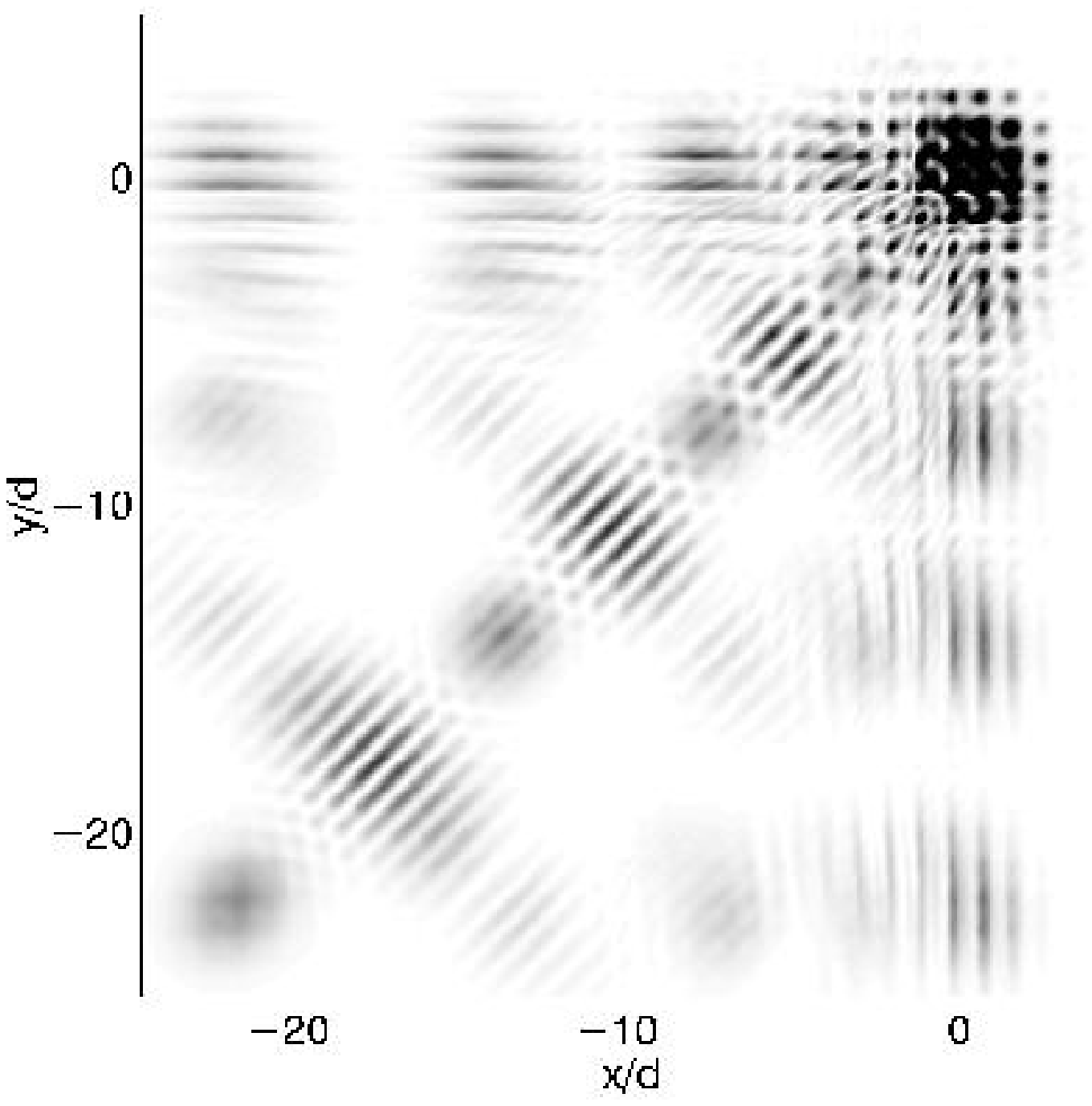}
\caption{A fragment of the wave function of the atom in the
separable (left panel) and non-separable (right panel)
potential in the case of a strong static force. The system
parameters are $\hbar=2$, $F=0.2$, and $F_x/F_y=1$.}
\label{fig_c5}   
\end{figure}

The case of a strong static force is essentially more complicated
and here the difference between the separable and non-separable
potentials appears on the {\em qualitative} level. Indeed, in the strong 
force regime, the atoms may escape out of the potential wells through a
sequence of Landau-Zener tunneling transitions to higher bands. 
The details of this process crucially depend on the particular
structure of the upper bands which, as seen in Fig.~\ref{fig6}, 
is quite different for $\varepsilon=0$ and $\varepsilon\ne0$. 
Similar to  the 1D case, one can analyze the tunneling of the
atoms by using the formalism of metastable (now 2D) Wannier-Stark
states. This analysis leads to the prediction
that in the separable case the atoms may escape 
out of the potential wells only along the $x$- and $y$-axis of the lattice,
while for the  non-separable potential additional escape channels
appear \cite{pra58}. This is illustrated in Fig.~\ref{fig_c5}, which
shows `snapshots' of two-dimensional BO in the strong force 
regime for $\varepsilon=0$ (left panel) and  $\varepsilon=1$ (right panel).
The additional chanel along the (1,1)-crystallographic direction is
clearly seen in the right panel. (Also notice a rich interference structure
between the channels, which is absent in the separable case.)

\subsection{Related problems}
\label{sec_c4}

The 2D optical lattice also offers an opportunity for studying
a number of related problems like, for example, `two-bands' BO. Indeed, 
superimposing the potentials shown in Fig.~\ref{fig_c1}, one can easily realize
the case, where the two lowest Bloch bands of the combined potential 
are separated by an arbitrary small gap (located along the edges
of the Brillouin zone). Then, performing a BO, the atom tunnels  
between these two bands at each crossing of the Brillouin zone, which 
results in a very non-trivial dynamics of BO \cite{pra58,Witt04}. 

The other problem we would like to mention is the scattering
of an atomic beam by a 2D optical potential \cite{pra55}. 
In classical dynamics, and for a non-separable potential, this would 
be a chaotic scattering process, with fractal basins for the 
scattered channels. The classical and quantum scattering of the
atoms by 2D lattices is studied in some details in the paper cited above.

\section{Decoherence of Bloch oscillations}
\label{sec_d0}

Since BO is a coherent quantum phenomenon, it is important to
study the processes which cause a decoherence of the system.
In this section we consider two of these processes -- 
decoherence due to spontaneous emission and due to
atom-atom interactions. We would like to stress that decoherence or 
relaxation, usually considered as `unwelcome' phenomena,
are also of interest on their own.
Indeed, as was already noticed by Esaki \cite{Esak74},
the conventional conductivity is an interplay between
BO and relaxation processes. Thus, the analysis of 
decoherence of BO is a necessary step for developing a theory 
of atomic conductivity.

\subsection{Decoherence by spontaneous emission}
\label{sec_d1}

As mentioned above in Sec.~\ref{sec_a2}, spontaneous
emission is a particular case of the systems interaction
with an environment. For this kind of problem, the approach based on
the single-particle Schr\"odinger equation is not applicable, and
the dynamics of the system should be described in terms of the density 
matrix $\rho(x,x';t)$, which is defined as the trace of the total 
wave function (system plus environment) over irrelevant 
variables of the environment. For a specified environment (the photon bath), 
the density matrix obeys the master equation \cite{Goet96},
\begin{equation}
\frac{\partial\hat{\rho}}{\partial t}=-\frac{i}{\hbar}[\widehat{H},\hat{\rho}]
+\frac{\tilde{\gamma}}{2}\int {\rm d}u P(u)
\left(\widehat{L}^\dagger_u \widehat{L}_u\hat{\rho}
-2\widehat{L}_u\hat{\rho} \widehat{L}^\dagger_u
+\hat{\rho} \widehat{L}^\dagger_u \widehat{L}_u\right) \;,
\label{d1}
\end{equation}
where $\widehat{H}$ is the Hamiltonian (\ref{b10}) (we consider
a quasi one-dimensional lattice), $\tilde{\gamma}$ the rate
of spontaneous emission (\ref{a7}), and $\widehat{L}_u$  the projection
of the recoil operator on the $x$ axis,
\begin{equation}
\widehat{L}_u=\cos(k_L x)\exp(iuk_L x) \;,\quad |u|\le 1 \;.
\label{d2}
\end{equation}
Note that  Eq.~(\ref{d1}) has the Lindblad form and, thus,
${\rm Tr}[\rho(t)]=\int dx \rho(x,x;t)=1$.
The distribution $P(u)$ of the random variable $u$ in Eq.~(\ref{d1}) is 
defined by the angle distribution for the momentum of the spontaneously 
emitted photons \cite{Leto81}, and is, in the case of linearly polarized 
light considered here, approximately given by $P(u)=1/2$.

\begin{figure}[!t]   
\center   
\includegraphics[width=12cm, clip]{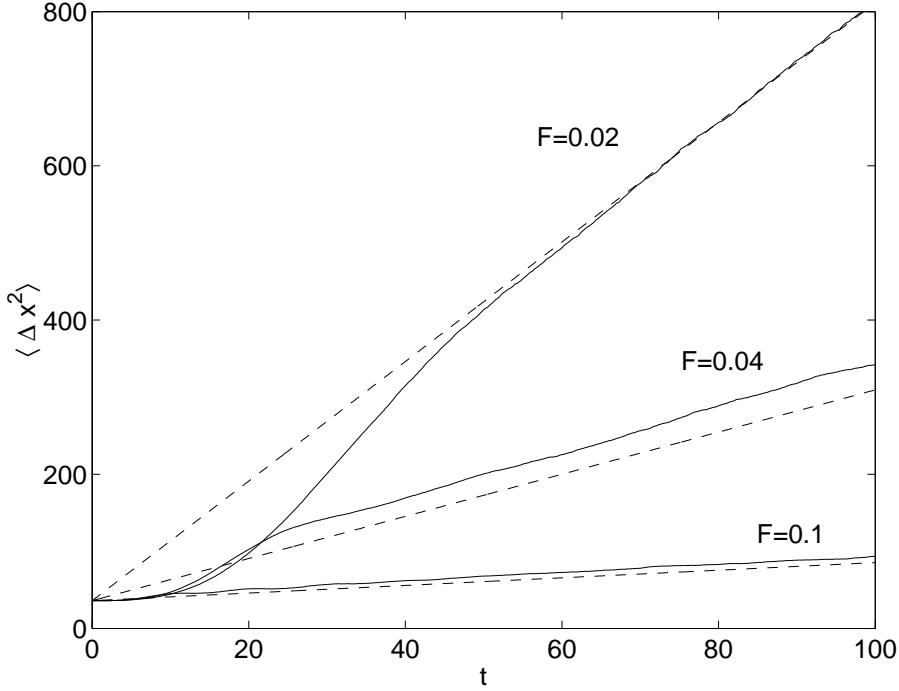}
\caption{Dispersion $\langle\Delta x^2\rangle=\langle x^2\rangle 
-\langle x\rangle^2$ of the atomic wave packet as a function of time.
Time is measured in the units of $T_J=2\pi\hbar/J$, 
the scaled rate of spontaneous emission
$\hbar\tilde{\gamma}/J=0.05$, $d=\pi$, and the scaled magnitude
of the static force ($F\rightarrow dF/J$) is indicated in the
figure. The slopes of the dashed lines are given by the
values of the diffusion coefficient (\ref{d3}).}
\label{fig8}   
\end{figure}

Using the tight-binding approximation, i.e.~substituting the
Hamiltonian $\widehat{H}$ by the tight-binding Hamiltonian (\ref{b17}), 
and the recoil operator (\ref{d2}) in its tight-binding version
$\widehat{L}_u=\sum_l (-1)^l\exp(i\pi u l)|l\rangle\langle l|$, 
the master equation (\ref{d1}) can be solved analytically, with 
the following main results \cite{pra56}: The spontaneous emission leads
to decoherence of the system, i.e.~the density matrix tends 
to a diagonal one in the basis of the Wannier states. 
As a consequence, BO decay as
\begin{equation}
\langle p(t)\rangle=p_0\exp(-\nu t)\sin(\omega_B t) 
\end{equation}
where the decay rate $\nu$ appears to coincide with
the rate $\tilde{\gamma}$ of the spontaneous emission.
The decay of BO is accompanied by (asymptotically)
diffusive spreading of the atoms (see Fig.~\ref{fig8}), 
\begin{equation}
\langle x^2(t)\rangle\sim D t \;,\quad D=\left(\frac{p_0}{M}\right)^2
\frac{\tilde{\gamma}}{\omega^2_B+\tilde{\gamma}^2} 
\label{d3}
\end{equation}
(here $p_0$ is the amplitude of BO in the absence
of the relaxation process). Note that (since $\omega_B\sim F$)
the static force actually suppresses the diffusion. Considering
the diffusion as a `generalized conductivity', this result
agrees with the prediction of Esaki and Tsu that the
conductivity of the system tends to zero when the 
frequency of BO becomes much larger than the characteristic frequency 
of the relaxation processes.

The (numerical) analysis of the system dynamics beyond the
tight-binding approximation leads to qualitatively the same
results, although the quantitative deviation can
be larger than 50 percent \cite{pra56}. It should also be mentioned
that the validity of Eq.~(\ref{d1}) still assumes a low population
of the excited electronic state of the atom and, hence, the 
case of resonant driving (briefly analyzed in Sec.~\ref{sec_b5}
for $\gamma=0$) is excluded from this consideration. 

\subsection{Interacting atoms and the Bose-Hubbard model}
\label{sec_d2}

Up to now, we have studied BO using single-particle 
quantum mechanics. This is justified only for a very dilute gas of
atoms, where the atom-atom interactions can be neglected.
If this is not the case, the problem
of BO becomes very diverse and, first of all,
one should distinguish between Bose and Fermi statistics.
In this review we restrict ourselves to bosonic atoms.
Moreover, we assume in what follows that the initial state
of the system is a Bose-Einstein condensate, where all atoms
occupy the zero quasimomentum state of the ground Bloch band.
To study the time evolution of this state, one usually uses 
the Gross-Pitaevskii (or nonlinear Schr\"odinger) equation.
It is understood, however, that this equation has a limited 
applicability -- in particular, the Gross-Pitaevskii equation
is unable to describe the decoherence of the system. Because of this, 
we employ here a more general approach, based on the Bose-Hubbard 
model. 

The Bose-Hubbard model (with a static term added) generalizes
the tight-binding Hamiltonian (\ref{b17}) to the multi-particle
case. To simplify the analysis we shall consider only the 1D 
Bose-Hubbard model, even when discussing the 3D lattices. (This
approximation is not crucial for the phenomena discussed below.) 
Then the Hamiltonian of the system has the form
\begin{equation}  
\widehat{H}= -
\frac{J}{2}\left(\sum_l \hat{a}^\dag_{l+1}\hat{a}_l +h.c.\right) 
+dF\sum_l l\hat{n}_l +\frac{W}{2}\sum_l\hat{n}_l(\hat{n_l}-1)  
\label{d4}  
\end{equation} 
(the band index $\alpha=0$ is omitted, $J>0$).
In the Hamiltonian (\ref{d4}), the creation operator $\hat{a}^\dag_l$
and the annihilation operator $\hat{a}_l$  `creates' or 
`annihilates' an atom in the $l$th well of the optical potential in the 
Wannier state $\psi_l(x)$. Hence, the first term on the
right hand side of Eq.~(\ref{d4}) is responsible for the tunneling
(hopping) of the atoms between the wells of the optical potential. 
Note that the hopping matrix elements $J$ (defined by the overlap
integral of the Wannier states in neighboring wells) 
exponentially depends on the depth of the optical potential and, hence,
can be easily varied by several order of magnitude
($0.30\le J/E_R\le0.0035$ for $2\le V_0/E_R\le22$).
The second term in the Hamiltonian (\ref{d4}) is the Stark energy
of the atoms in the homogeneous field. The third term is the
interaction energy of the atoms sharing one and the
same well, where the interaction constant $W$ is mainly defined by the
$s$-wave scattering length $a_{sc}$ of the atoms and by the
geometry of the lattice,
\begin{equation}  
\label{d5}
W=\frac{4\pi a_{sc}\hbar^2}{M}\int \psi^4_l({\bf r})d{\bf r} \;.
\end{equation}
As an estimate, one can use $W=0.28E_R$ -- the experimental
value for $^{87}$Rb atoms ($a_{sc}=5.8$ nm) in 3D separable
potential of the depth $V_0=22E_R$ \cite{Grei02b}. Interpolating this 
result to $V_0=2E_R$ would give $W=0.027E_R$.

A remark about the phase diagram of the Bose-Hubbard model
(the Hamiltonian (\ref{d4}) without the static term) is appropriate here.
As it is known, the Bose-Hubbard model shows the super-fluid/Mott-insulator 
quantum phase transition when the ratio between the parameters $J$ and 
$W$ exceeds some critical value \cite{Sach01}. 
(In the laboratory experiments one varies the ratio 
$W/J$ by changing the depth of the optical potential -- 
the critical value for $^{87}$Rb atoms is $V_0\approx15E_R$ and
$V_0\approx44E_R$ for the 3D and 1D lattices, respectively 
\cite{Grei02a,Orze01}.) Obviously, the dynamical response of the system
to a static force also depends on whether the system is
in the super-fluid or Mott-insulator regime. In the next section,
we study the Bloch dynamics in the super-fluid regime, where the
ground  state of the Bose-Hubbard system (which we take as the 
initial state in our numerical simulations) can be well approximated
by the product of Bloch waves with zero quasimomentum. 
The  response of the system to a static force in the Mott-insulator
regime will be briefly discussed in Sec.~\ref{sec_d5}.
\begin{figure}[!b]
\center
\includegraphics[width=12cm, clip]{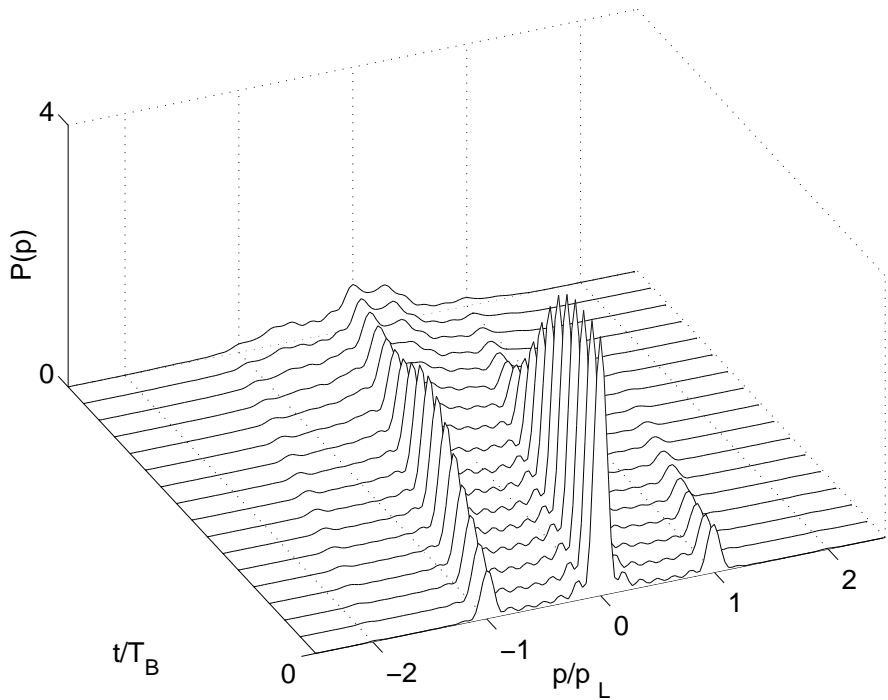}
\caption{Decay of BO due to the interaction induced decoherence.
Parameters are $J=0.038E_R$, $W=0.032E_R$, and $dF=-0.05E_R$.}
\label{fig9}
\end{figure}

\subsection{Decoherence due to the atom-atom interactions}
\label{sec_d4}

Numerically, the problem of BO of interacting bosonic atoms consists of
solving the Schr\"odinger equation for the multi-particle wave function 
$\Psi(t)=\sum_{\bf n} c_{\bf n}(t) |{\bf n}\rangle$, where
$|{\bf n}\rangle=|\ldots,n_{l-1},n_{l},n_{l+1},\ldots\rangle$
are the Fock (number) states, given by the symmetrized product
of Wannier states $\psi_l(x)$. Note that the translation
symmetry of the system, obviously broken by the static term,
can be actually recovered by using the gauge transformation,
\begin{equation}  
\widehat{H}\rightarrow\widehat{H}(t)= -
\frac{J}{2}\left(\sum_l e^{-i\omega_B t} \hat{a}^\dag_{l+1}\hat{a}_l
+h.c.\right) +\frac{W}{2}\sum_l\hat{n}_l(\hat{n_l}-1) \;. 
\label{d6}  
\end{equation}
This allows us to impose periodic boundary conditions, which
greatly facilitate the convergence in the thermodynamic limit
$N,L\rightarrow\infty$, $N/L=\bar{n}$. Here $L$ is the lattice
size and $N$ the number of atoms. For given $L$ and $N$, the
dimension of the Hilbert space (the total number of Fock states) is 
${\cal N}=(N+L-1)!/N!(L-1)!$. Note that the dimension of 
the Hilbert space but not the size of the system controls 
the convergence in the thermodynamic limit.

Having a (numerical) solution of the  Schr\"odinger equation,
$i\hbar\partial|\Psi(t)\rangle/\partial t=\widehat{H}(t)|\Psi(t)\rangle$, 
we then calculate the single-particle density matrix,
\begin{equation}  
\label{d7}
\rho(x,x';t)=\sum_{l,m}\psi_l(x)\psi_m(x')\rho_{l,m}(t) \;,\quad
\rho_{l,m}(t)=\langle\Psi(t)|\hat{a}^\dag_l\hat{a}_m|\Psi(t)\rangle \;,
\end{equation}
which carries essential (although not complete) information
about the system. In particular, the diagonal elements of the
density matrix (\ref{d7}) in the momentum representation define
the momentum distribution of the atoms $P(p)$ which, as discussed
above in Sec.~\ref{sec_a4}, is the quantity most easily measured in the
laboratory experiments.

\begin{figure}[!t]   
\center   
\includegraphics[width=12cm, clip]{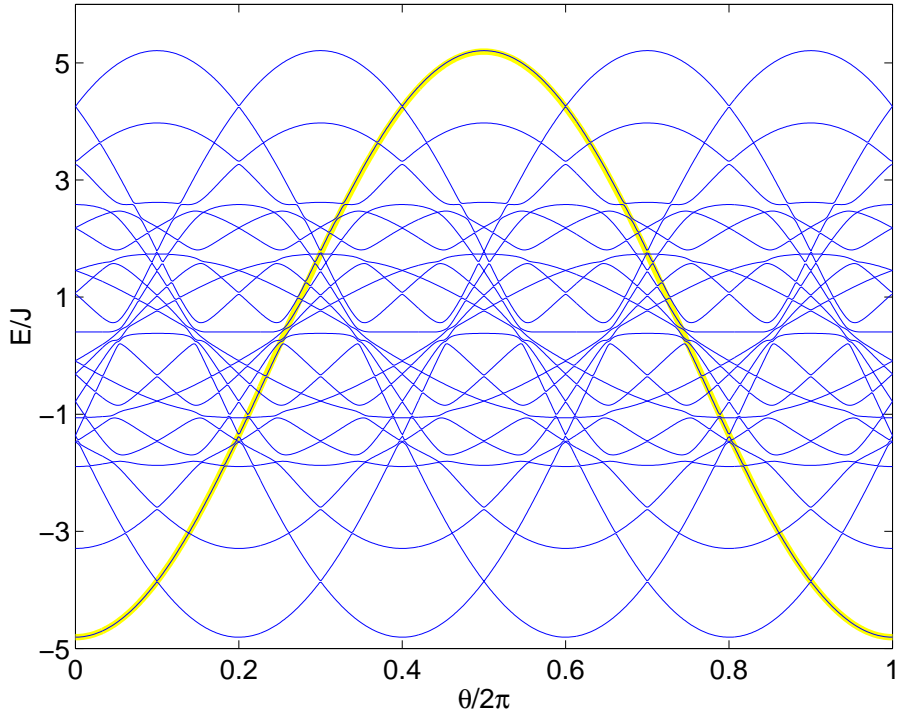}
\caption{Instantaneous spectrum of the Hamiltonian (\ref{d6}).
Lattice size $L=5$, number of the atoms $N=5$, interaction
constant $W=0.1J$.}
\label{fig10}   
\end{figure}

An example of the time evolution of the momentum distribution
is depicted in Fig.~\ref{fig9}. Comparing this figure with
Fig.~\ref{fig2} for BO of non-interacting atoms, a rapid decay of BO is 
noticed. The reason for this decay is the decoherence of the density
matrix, similar to that considered in Sec.~\ref{sec_d1}, 
but with the fundamental difference that here the system itself plays 
the role of a `bath'. To get a qualitative understanding 
of this phenomenon, it is useful to consider the instantaneous spectrum of
the time-dependent Hamiltonian (\ref{d6}) (see Fig.~\ref{fig10}). 
The thin line in Fig.~\ref{fig10} is the mean-field solution
(diabatic continuation of the ground state), where all atoms would 
oscillate coherently. However, due to Landau-Zener transitions at avoided 
crossings, the other Fock states become populated when the static
force drives the system along the the mean-field solution. This leads to 
a thermalization of the atoms and, as a consequence, to the decay of BO.

In principle, one can try to describe the thermalization process by
thoroughly analyzing the Landau-Zener tunneling at the avoided crossings.
An alternative (and, actually, more constructive) approach is to study 
the properties of the Floquet-Bloch operator \cite{pre60}, which we define
as the evolution operator over one Bloch period:
\begin{equation}  
\label{d7a}
\widehat{U}=\widehat{\exp}\left[-\frac{i}{\hbar}
\int_0^{T_B} \widehat{H}(t) dt\right] \;.
\end{equation}
It has been found that for the parameter region of a typical
experiment with cold atoms in the 3D lattices
($W\sim J$, $\bar{n}\sim 1$, and $dF<J$) the matrix of the
Floquet-Bloch operator (\ref{d7a}) can be well identified with a random matrix
of the circular orthogonal ensemble and, thus, the system (\ref{d4})
is a {\em quantum chaotic system} \cite{prl61,preprint2}. In fact, this is
precisely the Quantum Chaos, which justifies the use of the terms
`bath' and `thermalization' for the system of interacting atoms.
  
\subsection{Different regimes of BO of interacting atoms}
\label{sec_d5}

\begin{figure}[!t]   
\center   
\includegraphics[width=12cm, clip]{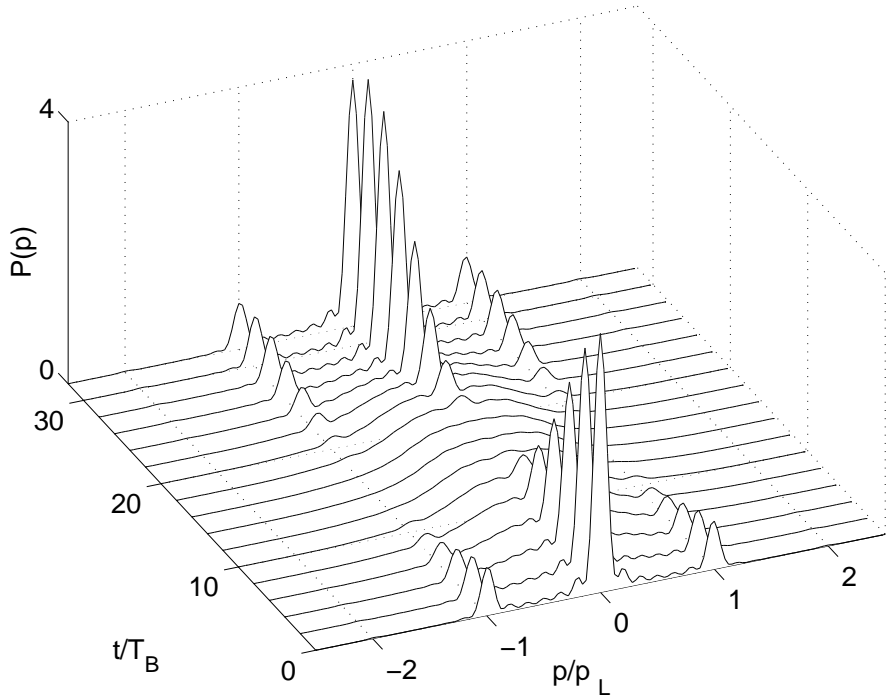}  
\caption{Quasiperiodic Bloch oscillations. Parameters $J$ and $W$
are the same as in Fig.~\protect\ref{fig9} but $dF=E_R$.
Note that the scale of the time axis is changed in comparison 
with Fig.~\protect\ref{fig9}.}
\label{fig11}   
\end{figure}

As already mentioned in Sec.~\ref{sec_d2}, BO of interacting atoms
is a rather diverse problem and the irreversible decay of BO discussed 
above is in no way the only possible regime of Bloch dynamics. In particular,
a static force with magnitude $dF\gg J$ suppresses the Quantum Chaos
and BO become quasiperiodic (see Fig.~\ref{fig11}).
This regime of BO can be treated analytically and, for example, 
for the mean atomic momentum  we have \cite{prl56},
\begin{equation}  
\label{d8}
\langle p(t)\rangle =p_0\sin(\omega_B t)\exp\{-2\bar{n}[1-\cos(\omega_W t)]\}
\;,\quad \omega_W=W/\hbar \;.
\end{equation}
Note, in passing, that the same interaction frequency 
$\omega_W=W/\hbar$ (but for a different problem) has recently been  
observed in a laboratory experiment \cite{Grei02b}.

The other limiting case of BO is the case of a weak atom-atom interactions,
which is typically realized in the quasi 1D lattices. Indeed, 
because of a weak transverse confinement $r_0\gg\lambda$,
the integral $\int \psi^4_l({\bf r})d{\bf r}$ in Eq.~(\ref{d5}) is
orders of magnitude smaller than
for 3D lattices. An increase of the filling factor 
from $\bar{n}\sim1$ in the 3D case to $\bar{n}\sim100$ in the 1D case 
(present days situation) may not be able to compensate this decrease of 
the interaction constant $W$ and, therefore, a regular dynamics of BO 
can be expected. Let us also remind that a large filling factor and a small 
interaction constant are usually considered as a validity condition for 
the Gross-Pitaevskii equation. We shall discuss this issue in more detail 
in Sec.~\ref{conclusion}.
 
It is also interesting to study `Bloch oscillations' in the Mott-insulator 
regime ($J\ll W$). In this case, a response of the system to a static force 
has a resonant character \cite{Grei02a,Brau02}, with the main resonance
corresponding to the condition $dF\approx W$. Providing resonant
(or near resonant) forcing, the particle-hole excitations of the
Mott-insulator states are created dynamically. This excitation process
is reflected in the (almost) periodic dynamics of the atomic momentum
distribution \cite{preprint1}. It should be stressed, however, that
in spite of the formal analogy with BO, the origin of these oscillations 
as well as the characteristic period $T_J\sim\hbar/J$ are fundamentally 
different.

\section{Conclusion}
\label{conclusion}

We considered the phenomenon of BO for cold neutral atoms
in optical lattices. In Sec.~\ref{sec_a0} and 
Sec.~\ref{sec_b0} of the review, which mainly serve as
tutorials, we focused on a dilute gas of atoms in quasi
1D lattices. In addition to the theoretical analysis, we also
discussed the scheme of a typical laboratory experiment on BO
and indicated the characteristic values of the parameters.
 
It can be safely  stated that in the above case of a dilute gas in 
1D lattices the BO of cold atoms are well understood. 
The further progress in the field is related to the problems of
BO in lattices of higher dimensionality, BO in the presence
of relaxation processes, and BO of interacting atoms.

A dilute gas of cold atoms in 2D lattices was discussed in 
Sec.~\ref{sec_c0}. Naively, one may expect the two-dimensional BO to be  
a superposition of one-dimensional BO, where the values of the static
force $F_{x,y}$ are given by the projection of the 
vector ${\bf F}$ to the crystallographic axis of the lattice.
However, this is true only in the separable case. For a non-separable
optical potential, the two-dimensional BO generally cannot be 
expressed in terms of the 1D problem. This is especially the case for 
a strong static force (strong Landau-Zener tunneling), where 
non-separability of the potential manifests itself in a number of effects.

The ultimate goal for studying BO in the presence of 
relaxation processes is to obtain a directed diffusive current
of atoms, similar to that of electrons in a conductor.
In Sec.~\ref{sec_c1} we studied the relaxation (decoherence)
of BO due to spontaneous emission. Note, that the rate
of spontaneous emission can be increased to any desired level
by simply tuning the laser frequency closer to the atomic resonance.
The effect of spontaneous emission is shown to lead only to 
`non-directed' diffusion. The theoretical analysis of the other relaxation 
mechanisms, like scattering on `impurities' is strongly in need.
In this connection we would like to mention the recent     
experiment \cite{Ott03}, which studies the dynamics of the fermionic $^{40}$K
atoms with an admixture of bosonic $^{87}$Rb atoms in inhomogeneous
optical lattices.

Finally, we comment on the problem of BO for a Bose-Einstein condensate.
This field of research opens unlimited perspectives for studying
the different phenomena of correlated systems. 
In particular, by loading a BEC of cold atoms into a quasi 1D lattice,
one can realize the case of a large filling factor (mean number of atoms 
per lattice cite) and a small interaction constant. This is believed to be
the realm of the Gross-Pitaevskii equation, where BO becomes a
macroscopic quantum phenomenon. We intentionally did not discuss this
case because even the mentioning of all publications on this subject  
would draw this review out of the length limit. Instead, we analyzed
BO of interacting atoms in the 3D lattice. Loading BEC 
in the 3D lattice increases the interaction constant by several orders
of magnitude and simultaneously decreases the filling factor to 
${\bar n}\sim1$. In this regime the Gross-Pitaevskii equation fails to  
describe the dynamics of the atoms and a fully microscopic treatment 
of the system is required. Such a microscopic approach is provided by the
Bose-Hubbard model. It was shown that, depending on the value
of the static force, BO of interacting atoms may vary from quasiperiodic 
to irreversibly decaying behavior. Moreover, in the latter case, the 
system appears to be chaotic in the sense of Quantum Chaos. This 
intriguing regime of BO is waiting for experimental studies.

\vspace*{2mm}
Financial support by Deutsche Forschungsgemeinschaft 
(Grand No. 436 RUS 113/658/0-1) is acknowledged. We also
express our gratitude to A.~Buchleitner, who is our co-author
in Sec.~\ref{sec_d0} of the present review.


\end{document}